\documentclass{emulateapj}
\usepackage{apjfonts}

\newcommand{\p}{\left[{\cal P}_3\right]}
\newcommand{\pp}{\left[{\cal P}_4\right]}

\shorttitle{Black Hole Spin Dependence}
\shortauthors{Barai, Das \& Wiita}

\begin{document}


\title{Dependence of General Relativistic Accretion on Black Hole Spin}

\author{Paramita Barai$^{1}$, Tapas K. Das$^{2}$, Paul J. Wiita$^{1}$}
\affil{$^1$ Department of Physics \& Astronomy, Georgia State University,
P.O. Box 4106, Atlanta GA 30302-4106 \\
Email: barai@chara.gsu.edu, wiita@chara.gsu.edu\\
$^2$ N. Copernicus Astronomical Center, Bartycka 18, Warsaw 00-716 Poland\\
Email: tapas@camk.edu.pl}



\begin{abstract}
{We propose a novel approach to the investigation of how black hole spin influences
the behavior of astrophysical accretion. We examine the terminal behavior
 of general relativistic, multi-transonic,
hydrodynamic advective accretion in the Kerr metric, by analyzing various 
dynamical as well as thermodynamic
properties of accreting matter extremely close to the event horizon, as a function of
Kerr parameter $a$.  The examination of the transonic properties of such flows can be  
useful in self-consistently discriminating between prograde and retrograde 
relativistic accretion, and in studying the spectral signatures of black hole rotation.
}
\end{abstract}


\keywords{accretion, accretion disks --- black hole physics --- hydrodynamics
--- relativity }


\section{Introduction}
\noindent
General relativistic (GR) accretion flows dive supersonically onto astrophysical
black holes (BH) in order to satisfy the inner boundary condition imposed by the event
horizon. Hence such accretion flows must be globally transonic (e.g.,
Fukue 1987; Chakrabarti 1989).
This extremely hot, high density, ultra-fast
flow can provide 
key features of the  spectra of some galactic and extragalactic
BH candidates. 
It is 
believed that most of the supermassive BHs powering active 
galactic nuclei (AGN) and stellar mass BHs in galactic microquasars 
possess non-zero values of the BH spin parameter, $a$,
which might play a crucial role in determining their spectra and jet production
(e.g., Rees 1984; Frank, King \& Raine 1992; Kato, Fukue \& Mineshige 1998). 
Hence it is very important to study the
dynamical as well as the thermodynamic properties of GR transonic 
BH accretion very
close to the event horizon and to investigate how BH spin modulates such 
behavior. 

Study of the multi-transonic behavior of BH accretion was initiated by
Abramowicz \& Zurek (1981).
Here we formulate and solve
the conservation equations governing GR,  multi-transonic, advective 
accretion in the Kerr metric to calculate the flow behavior extremely close 
($\simeq$ 0.01 gravitational radii) to the BH event horizon as a function 
of the Kerr parameter, $a$. 
Such a solution scheme
for accretion onto a non-rotating BH has been outlined in Das (2004).
We concentrate on accretion with sub-Keplerian angular
momentum distributions.
Such weakly rotating flows
are exhibited in
various physical situations, such as detached binary systems
fed by accretion from OB stellar winds (Illarionov \&
Sunyaev 1975; Liang \& Nolan 1984), semi-detached low-mass
non-magnetic binaries (Bisikalo et al.\ 1998), and super-massive BHs fed
by accretion from slowly rotating central stellar clusters
(Illarionov 1988; Ho 1999 and references therein). Even for a standard Keplerian
accretion disk, turbulence may produce such low angular momentum flow 
(e.g., Igumenshchev
\& Abramowicz 1999, and references therein).
We find that our formalism is also valid for accretion with substantial
angular momentum (as large as Keplerian) for retrograde flows.

\section {Formalism}
\noindent
We consider non-self-gravitating, non-magnetized accretion. The gravitational radius is
$r_g= G M_{BH}/c^2$ and
we employ units where $G=c=M_{BH}=1$.
We use Boyer--Lindquist
coordinates with signature $-+++$, and an
azimuthally Lorentz boosted orthonormal tetrad basis co-rotating
with the accreting fluid. We define $\lambda$ to be the specific
angular momentum of the flow and neglect any gravo-magneto-viscous
non-alignment between $\lambda$ and $a$.
We focus on the stationary
axisymmetric solution of the following equations:

\begin{equation}
{\bf \nabla}^\mu{\Im}^{{\mu}{\nu}}
=0,~
~
\left({\rho}{v^\mu}\right)_{;\mu}=0,
\end{equation}

\noindent 
where ${\Im}^{{\mu}{\nu}}, v_\mu$, and $\rho$ are the energy momentum tensor, the four
velocity and the rest mass density of the 
accreting fluid, respectively, and the semicolon  denotes the 
covariant derivative. The radial momentum balance condition
may be obtained from
$\left(v_{\mu}v_{\nu}+g_{{\mu}{\nu}}\right){\Im}^{{\mu}{\nu}}_{;{\nu}} = 0$.
Exact solutions of these conservation equations
require knowledge of the accretion
geometry and the introduction of a suitable equation of
state. 
We concentrate on polytropic accretion for which
$p=K{\rho}^{\gamma}$, where $K$ and $\gamma$
are the monotonic and continuous
functions of the specific entropy density, and  the constant
adiabatic index of the flow, respectively.
The specific proper flow enthalpy is taken to be
$h=\left(\gamma-1\right)\left\{{\gamma}-\left(1+a_s^2\right)\right\}^{-1}$,
where $a_s$ is the polytropic sound speed defined as 
$a_s=\left(\partial{p}/\partial{\epsilon}\right)_{\cal S}^{1/2}
={\bf \Psi_1}\left(T(r),\gamma\right)={\bf \Psi_2}\left(p,{\rho},\gamma\right)$;
here $T(r)$ is the local flow temperature,
$\epsilon$ is the mass-energy density,
and $\left\{{\bf \Psi_1},{\bf \Psi_2}\right\}$ are known
functions. The subscript ${\cal S}$ indicates that the derivative
is taken at constant specific entropy. Hence, 
$a_s^2(r)= \gamma{\kappa}T(r)/({\mu}m_H)={\Theta}^2T(r)$,
where $\Theta = [\gamma{\kappa}/(\mu{m_H})]^{1/2}$,
$\mu$ is the mean molecular weight,
$m_H$ is the mass of the hydrogen atom,
and $\kappa$ is Boltzmann's constant. 

We assume that the disk has a radius dependent local
half-thickness, $H(r)$, and its central plane coincides with
the equatorial plane of the BH.
We use the vertically integrated
model (Matsumoto et al.\ 1984) in
describing black hole accretion disks where the equations of motion
apply to the equatorial plane of the BH, assuming the flow to
be in hydrostatic equilibrium in the transverse direction. 
Various dissipative processes are not explicitly
taken into account. 
Even thirty years after the discovery of standard accretion disk theory
(Shakura \& Sunyaev 1973), modeling of viscous multi-transonic
BH accretion, including proper heating and cooling mechanisms, is still
quite an arduous task, even for Newtonian flows. 
The extremely large radial velocity
close to the BH implies $\tau_{in} \ll \tau_{vis}$, with $\tau_{in} (\tau_{vis})$ 
the infall (viscous) timescales; hence
our assumption of inviscid flow is not unjustified, at least out to
a few 
or to perhaps tens of $r_g$. 
One of the most significant effects of the introduction of
viscosity would be the reduction of radial angular momentum.
Hence study of the behavior of flow variables as functions of
$\lambda$ provides some insight about the flow behavior for
viscous transonic accretion.

The temporal component of the first part of Eq.\ (1) leads to
the conservation of specific flow energy ${\cal E}$
along each streamline
as,  ${\cal E}=hv_t$ (Anderson 1989).
The Kerr metric in the equatorial
plane of the black hole may be written 
as (e.g., Novikov \& Thorne 1973)

\begin{equation}
ds^2=g_{{\mu}{\nu}}dx^{\mu}dx^{\nu}=-\frac{r^2{\Delta}}{A}dt^2
+\frac{A}{r^2}\left(d\phi-\omega{dt}\right)^2
+\frac{r^2}{\Delta}dr^2+dz^2 ,
\end{equation}

\noindent
where
$\Delta=r^2-2r+a^2, ~A=r^4+r^2a^2+2ra^2$, 
and $\omega=2ar/A$.
The angular velocity $\Omega$ is

\begin{equation}
{\Omega}=-\frac{\left(g_{t\phi}+\lambda g_{tt}\right)}
{\left(g_{{\phi}\phi}+\lambda g_{t\phi}\right)}
=\frac{\frac{4a}{r}-\frac{\lambda \left(4a^2-r^2\Delta \right)}{A}} {\frac{A}{r^2}-
\frac{4{\lambda}a}{r}}.
\end{equation}

The normalization relation $v_\mu{v^\mu}=-1$, along with the above 
value of $\Omega$,  
gives us the expression

\begin{equation}
v_t=
\left[\frac{Ar^2\Delta}
{\left(1-u^2\right)\left\{A^2-8\lambda arA
+\lambda^2r^2\left(4a^2-r^2\Delta\right)\right\}}\right]^\frac{1}{2},
\end{equation}

\noindent
where $u$ is the radial three velocity in the co-rotating fluid frame.
Hence we obtain
the conserved specific energy (which includes the rest mass energy) to be
${\cal E}=\left(\gamma-1\right) \left[\gamma-\left(1+\Theta^2T\right)\right]^{-1}
v_t.$


Note that our final calculations will provide the value of $u$ and its
related quantities rather than those of $v_t$.  
Hereafter, it is implied that any appearance of $v_t$
in any mathematical expression is transformed in terms of $u$
using Eq. 
(4).

We follow  Abramowicz, Lanza \& Percival (1997) 
to derive the expression for disk
height $H(r)$ in our flow geometry because 
the corresponding equations in their 
calculations remain non-singular on the horizon and can 
accommodate a thin disk geometry 
as well as quasi-spherical flow.
We obtain:

\begin{equation}
H(r)=\sqrt{2}r^2\Theta T^{\frac{1}{2}}
\left[\left(\frac{\gamma-1}{\gamma}\right)
\left(\frac{1}{\gamma-\left(1+\Theta^2T\right)}\right)
\left(\frac{1}{\psi}\right)\right]^{\frac{1}{2}},
\end{equation}

\noindent
where $\psi = \lambda^2v_{t}^2-a^2\left(v_{t}-1\right)$. We define the 
quasi-invariant entropy accretion 
rate, ${\dot {\bf \Xi}}$, as a quasi-constant
multiple of the mass accretion rate,
$
{\dot {\bf \Xi}}
=K^{\frac{1}{\gamma-1}}{\dot M}_{in},
$
where $d\dot{\bf \Xi}/dr=0$
along a streamline 
only for a shock-free non-dissipative flow,
and not for a flow containing shocks. The mass accretion rate ($\dot M$) and the 
entropy accretion rate are found to be

\begin{equation}
{\dot M}=4{\pi}{\rho}{\Theta}T^{\frac{1}{2}}Mr^2
\left[\frac{2\Delta\left(\gamma-1\right)\Theta^2T}
           {\gamma \left(1-\Theta^2TM^2\right) \left\{\gamma-\left(1+\Theta^2T\right)\right\}
            \psi}\right]^{\frac{1}{2}};
\end{equation}


\begin{equation}
{\dot {\bf \Xi}}=
4\sqrt{2}\pi r^2 u
\sqrt{ \frac{\Delta} {\left(1-u^2\right)\psi} }
\left[\frac{a_{s}^2\left(\gamma-1\right)}{\gamma\left\{\gamma-\left(1+a_{s}^2\right)\right\}}\right]
^{{\frac{1}{2}}\left(\frac{\gamma+1}{\gamma-1}\right)}.
\end{equation}

We simultaneously solve eqs. (4 -- 7) along with the equation defining
${\cal E}$, to find the dynamical three-velocity gradient as

\begin{equation}
\frac{du}{dr}=
\frac
{ \frac{2a_{s}^2}{\left(\gamma+1\right)}
  \left[ \frac{r-1}{\Delta} + \frac{2}{r} -
         \frac{v_{t}\sigma \chi}{4\psi}
  \right] -
  \frac{\chi}{2} }
{ \frac{u}{\left(1-u^2\right)} -
  \frac{2a_{s}^2}{ \left(\gamma+1\right) \left(1-u^2\right) u }
   \left[ 1-\frac{u^2v_{t}\sigma}{2\psi} \right] },
\end{equation}

\noindent where

\begin{equation}
\sigma = 2\lambda^2v_{t}-a^2,
~
\chi =
\frac{1}{\Delta} \frac{d\Delta}{dr} +
\frac{\lambda}{\left(1-\Omega \lambda\right)} \frac{d\Omega}{dr} -
\frac{\left( \frac{dg_{\phi \phi}}{dr} + \lambda \frac{dg_{t\phi}}{dr} \right)}
     {\left( g_{\phi \phi} + \lambda g_{t\phi} \right)}.
\end{equation}

Hereafter, we use the notation $\left[{\cal P}_3\right]$ for a set
of values of $\left\{{\cal E},\lambda,\gamma\right\}$ for a fixed
Kerr parameter $a$ (including the counter-rotating flows where
$a$ and $\lambda$ have opposite signs), 
and $\left[{\cal P}_4\right]$ for a set of values
of $\left\{{\cal E},\lambda,\gamma,a\right\}$.
By simultaneously setting
the numerator and denominator of Eq.\ 
(8) to zero 
we find the sonic point conditions:

\begin{equation}
a_{s}{\vert}_{c}={\left[\frac{u^2\left(\gamma+1\right)\psi}
                                  {2\psi-u^2v_t\sigma}
                       \right]^{\frac{1}{2}}_{c}  },
~~u_c= {\left[\frac{\chi\Delta r} {2r\left(r-1\right)+ 4\Delta} \right]
^{\frac{1}{2}}_{c}  },
\end{equation}

\noindent
where the subscript $c$ indicates that the quantities are to be 
measured
at the critical (sonic) point(s). For a fixed $\left[{\cal P}_3\right]$ and $a$,
we substitute the values of $u_c$ and $a_{s}{\vert}_{c}$  in terms of $r_c$ 
(Eq.\ 10)
in the expression 
for ${\cal E}$,
and obtain
a polynomial in $r_c$, the solution of which provides
the location of the sonic point(s) $r_c$.
To
determine the behavior of the solution near the sonic
point, one needs to evaluate the value of $du/dr$
at that
point (the `critical velocity gradient' $(du/dr)_c$)
by applying L'Hospital's rule to Eq.\ (8).
After lengthy algebraic manipulations,
we obtain the following quadratic equation 
which can be solved 
to obtain $(du/dr)_c$

\begin{equation}
\alpha \left(\frac{du}{dr}\right)_c^2 + \beta \left(\frac{du}{dr}\right)_c + \zeta = 0,
\end{equation}
\noindent
where the coefficients are 


\begin{displaymath}
\alpha=\frac{\left(1+u^2\right)}{\left(1-u^2\right)^2} - \frac{2\delta_1\delta_5}{\gamma+1},
~~~\beta=\frac{2\delta_1\delta_6}{\gamma+1} + \tau_6,
~~~\zeta=-\tau_5;
\end{displaymath}
\begin{displaymath}
\delta_1=\frac{a_s^2\left(1-\delta_2\right)}{u\left(1-u^2\right)},
~\delta_2 = \frac{u^2 v_t \sigma}{2\psi},
~\delta_3 = \frac{1}{v_t} + \frac{2\lambda^2}{\sigma} - \frac{\sigma}{\psi} ,
\end{displaymath}
\begin{displaymath}
\delta_4 = \delta_2\left[\frac{2}{u}+\frac{u v_t \delta_3}{1-u^2}\right],
~
\delta_5 = \frac{3u^2-1}{u\left(1-u^2\right)} - \frac{\delta_4}{1-\delta_2} -
           \frac{u\left(\gamma-1-a_s^2\right)}{a_s^2\left(1-u^2\right)},
\end{displaymath}
\begin{displaymath}
\delta_6 = \frac{\left(\gamma-1-a_s^2\right)\chi}{2a_s^2} +
           \frac{\delta_2\delta_3 \chi v_t}{2\left(1-\delta_2\right)},
\end{displaymath}
\begin{displaymath}
\tau_1=\frac{r-1}{\Delta} + \frac{2}{r} - \frac{\sigma v_t\chi} {4\psi},
~
\tau_2=\frac{\left(4\lambda^2v_t-a^2\right)\psi - v_t\sigma^2} {\sigma \psi},
\end{displaymath}
\begin{displaymath}
\tau_3=\frac{\sigma \tau_2 \chi} {4\psi},
~
\tau_4 = \frac{1}{\Delta} 
       - \frac{2\left(r-1\right)^2}{\Delta^2}
       -\frac{2}{r^2} - \frac{v_t\sigma}{4\psi}\frac{d\chi}{dr},
\end{displaymath}
\begin{displaymath}
\tau_5=\frac{2}{\gamma+1}\left[a_s^2\tau_4 -
     \left\{\left(\gamma-1-a_s^2\right)\tau_1+v_ta_s^2\tau_3\right\}\frac{\chi}{2}\right]
   - \frac{1}{2}\frac{d\chi}{dr},
\end{displaymath}
\begin{equation}
\tau_6=\frac{2 v_t u}{\left(\gamma+1\right)\left(1-u^2\right)}
       \left[\frac{\tau_1}{v_t}\left(\gamma-1-a_s^2\right) + a_s^2\tau_3\right].
\end{equation}
Note that all the above quantities are evaluated at $r_c$.\\

\begin{figure}
\plotone{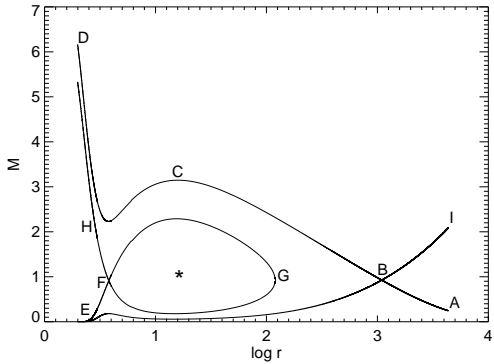}
\caption{A typical flow topology for multi-transonic BH accretion: 
local Mach number plotted against distance from the event horizon of the  BH,
for $a = 0.2$, ${\cal E} = 1.0003$, $\lambda = 3.1$, and $\gamma = 4/3$.
The curve {\bf ABCD}
represents accretion passing through the outer sonic point {\bf B} ($r_o = 1094$),
while
{\bf EBI} represents the self-wind. Flow along {\bf GFH} passes
through the inner sonic point {\bf F} ($r_i = 3.796$) and
encompasses a middle sonic point $r_{m}$ ($r_{m} = 14.78$) denoted
by an asterisk.  A multi-transonic accretion flow passes through {\bf AB}, then
undergoes a shock which causes it to drop to  the lower branch of the
{\bf GFH} curve.
}
\end{figure}

\begin{figure}
\plotone{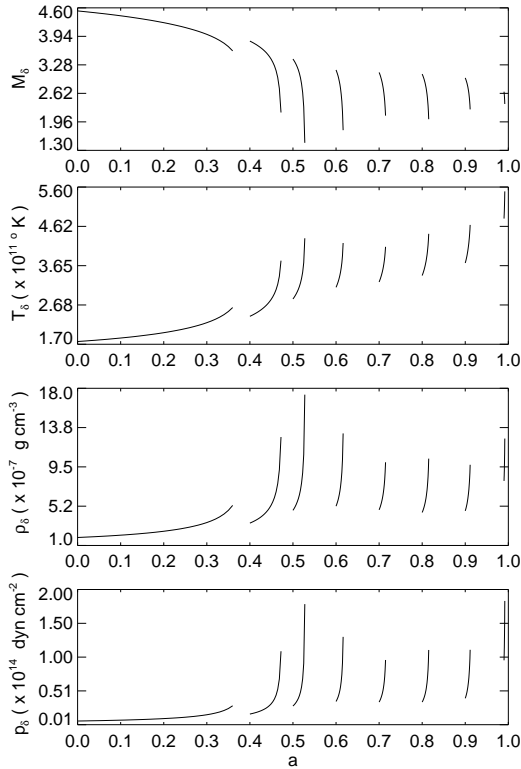}
\caption{ Variation of terminal behavior of accretion flow variables,
${\rm M}_{\delta},  T_{\delta}, \rho_{\delta}$, and $p_{\delta}$
as functions of BH spin, $a$. The figure is drawn for a shock-free 
prograde flow passing through the outer sonic point, with ${\cal E}$ = 1.00001;
the curves (from left to right) correspond to values 
of $\lambda$ = 2.60, 2.12, 1.87, 1.55, 1.27, 1.04, 0.829, and 0.609. 
See text for details.
}
\end{figure}

\section{Results}
For a range of $\pp$, one obtains three different values of $r_c$ for a fixed $\pp$,
among which the largest and the smallest values correspond to the $X$ type outer, $r_o$, and
inner, $r_i$, sonic points, respectively. The $O$ type middle sonic point, $r_m$, which is
unphysical in the sense that no steady transonic solution passes through it, lies
between $r_i$ and $r_o$. For $\Xi(r_i)>\Xi(r_o)$ and $\Xi(r_i)<\Xi(r_o)$, one obtains 
multi-transonic accretion and a wind, respectively.
If the accretion through $r_o$ can be perturbed in a
way so that it produces an amount of entropy exactly equal
to $\left[\Xi(r_i)-\Xi(r_o)\right]$, 
supersonic flow through $r_o$ can join the subsonic flow through $r_i$ by developing 
a standing shock. The exact location of such a shock, as well as the details of the post-shock
flow, may be obtained by formulating and solving the general relativistic Rankine-Hugoniot
condition (Taub 1948; Thorne 1973) for our flow geometry.
It is worth noting that the idea of shock formation in BH accretion
is contested by some authors
(see Narayan et al.\ 1998 and references therein for a review).
However, the problem of not finding shocks in such works
perhaps lies in the fact that
only one (inner) sonic point close to the BH
is explored by shock-free ADAF solutions.

For any $\pp$, after obtaining $\left[r_i,r_m,r_o\right]$, we calculate the values of
dynamical and acoustic velocities 
(Eq.\ 10) and the velocity gradient 
(Eq.\ 11)
for $\left[r_i,r_o\right]$.
These values are evaluated
for a specific value of ${\dot M}$ (= 1.0 ${\dot M}_{Edd}$, in the results
presented here; however, ${\dot M}$ could be chosen
greater or less than the Eddington rate), and we can then compute 
$u(r)$, $a_s(r)$, the local Mach number,
${\rm M}(r)$, and $T(r)$.  Here we define ${\rm M}(r) = u(r)/a_s(r)$, but 
because of our assumption of vertical equilibrium in an integrated
disk model, ${\rm M}(r_c) < 1$ (Matsumoto et al.\ 1984; Chakrabarti 1989). 
We compute the above quantities, 
the local vertically integrated density, ${\Sigma}(r)$, and pressure, $W(r)$ 
(Matsumoto et al.\ 1984), 
and other related dynamical or thermodynamic quantities by
integrating the flow equations, from 
$r_o$ and $r_i$
down to the immediate vicinity of the event horizon,
using a fourth order Runge-Kutta method. The flow could also 
be integrated for $r>r_c$ to study
the subsonic part of the flow in detail. Figure 1 shows the integral curves of motion for 
a particular case. 
Similar topologies are obtained for any $\pp$ allowing multi-transonic
accretion flows. 
We find that $r_{i,pro} < r_{i,retro}$ for all $\p$ for which valid common solutions
are available for both $a$ and $-a$; however, for $a \gtrsim 0.2$, no 
such common $\p$ solutions can
be found.

We define the ``terminal'' value of any accretion variable $A$ at a radial distance
$r_\delta=\left(r_e+{\delta}\right)$ as $A_\delta$, where $r_e$ is the event 
horizon, and $0<\delta \ll 1.0$. 
Here we calculate $A_\delta$ by integrating the 
flow from $r_c$ down to $r_\delta$, so is clear that the dependence of $A_\delta$
on $a$ can be studied at {\it any} $r$. For $\delta=0.01$ and $M_{BH} = 10 M_{\odot}$,
Fig.\ 2 shows the variations of
$M_\delta, T_\delta, \rho_\delta$ and $p_\delta$ with BH spin for shock free 
prograde flows passing through $r_o$. 

Note that as a flow through $r_i$ does not connect
$r_e$ with infinity (such a flow folds back onto itself, see Fig.\ 1), solutions through
$r_i$ do not have {\it independent} physical existence, and can only be accessed 
if the supersonic flow through $r_o$ undergoes a shock, so that it generates extra
entropy, becomes subsonic, and produces the physical segment of the solution through $r_i$.
If a shock forms, the calculation of $A_\delta$ for shocked flows boils down to the
calculation of $A_\delta$ for flow through $r_i$. 
We find that for flows through $r_i$, the general profiles for the
variations of $A_\delta$ with $a$ remain exactly the same as that for
flow through $r_o$; only the magnitudes of various $A_\delta$ change.
We obtain ${\rm M}_\delta(r_i)< {\rm M}_\delta(r_o)$ and
$\left[T_\delta,\rho_\delta,p_\delta\right](r_i)>
\left[T_\delta,\rho_\delta,p_\delta\right](r_o)$, 
because the shock compresses the flow and makes it hotter.
Fig.\ 2 is drawn for $\gamma=1.43$, but a similar figure could be drawn for any value of
$\gamma$ allowing multi-transonic accretion. 
We find that for both prograde and retrograde flows, 
$\left[T_\delta,\rho_\delta,p_\delta\right]$ non-linearly and monotonically
correlate with $\gamma$ while ${\rm M}_\delta$ anti-correlates with $\gamma$.
Multi-transonic accretion does not exist for the same $\p$ for all values of $a$ 
for prograde flows. Each curve segment in Fig.\ 2 shows the range of $a$ for 
which a set of $\left\{{\cal E},{\lambda}\right\}$ overlaps to produce three sonic points. 
Specifically, ${\cal E}$ is constant in Fig.\ 2 and each curve corresponds
to a value of $\lambda$ which allows
multi-transonicity for that range of $a$. The allowed overlap range 
in $a$
decreases as $a$ increases.

For retrograde flows, the situation is  different. For many combinations of
$\p$, one obtains a  large range of $a$ 
(sometimes even covering the {\it complete} range of BH spin) 
for which one can have a {\it fixed} 
value of $\left\{{\cal E}, \lambda, {\gamma}\right\}$ producing multi-transonicity.
Hence the variations of $A_\delta$'s with $a$ produce continuous curves covering almost
the whole range $-1 < a \le 0$. {\it Therefore multi-transonicity is
more common for counter-rotating BH accretion.} We also found that, 
unlike 
prograde flows, retrograde flows with
values of angular momentum even as high as Keplerian can demonstrate multi-transonic
behavior.

From Fig.\ 2 we see that the 
the correlations (of temperature, density and pressure) and the 
anti-correlation (of Mach number), 
with BH spin are non-linear but monotonic. 
Hence, rapidly rotating BHs will
produce a hotter and denser {\it prograde} flow near the horizon. 
However, we again found that
this is not true for counter-rotating flows, where the 
above-mentioned variations are not
monotonic; rather they show distinct points of inflection at 
some intermediate value of $a$.
The physical reason for such a behavior for retrograde 
flows is not clear to us.
 If one keeps $\left\{{\cal E},\lambda,|a|\right\}$ fixed
but decreases $\lambda$
(which effectively mimics the introduction of viscosity to the system),
$\left[T_\delta,\rho_\delta,p_\delta\right]$ non-linearly decreases and ${\rm M}_\delta$ 
increases for prograde flow, whereas the trends are just the opposite for retrograde flow.
These trends indicate how our results would be modified if our formalism were to be 
applied to study viscous transonic flow.
Hence the study of $A_\delta$ as a function of $a$ also opens up the possibility of 
distinguishing between the prograde and retrograde flows if the transonic
properties of such flows can be measured.

\section{Discussion}
Our formalism clearly allows the
study of the dependence of the behavior of multi-transonic
accretion on BH spin at any accretion length scale (including the radial shock location).
Thus our work is  important for understanding how BH spin influences
several accretion related astrophysical phenomena, as only multi-transonic 
flows can produce shock waves
in BH accretion disks (Das 2002; Das, Pendharkar \& Mitra 2003, and references therein).
Such hot and dense post-shock flow
is important for
understanding astrophysical phenomena such as
spectral properties of BH candidates (Chakrabarti \& Titarchuk 1995, and
references therein),
the formation and dynamics of accretion powered cosmic
jets (Das \& Chakrabarti 1999, and references
therein; Das, Rao \&
Vadawale 2003), and the origin of QPOs in
galactic sources (Das 2003, and references therein).
Once we compute the exact shock location and related post-shock
quantities 
(Barai, Das \& Wiita, in preparation)
for flow geometries presented in this Letter, it becomes straightforward 
to show how $a$ determines the exact macrophysics of the post-shock flow right
at the shock location, which in turn, may help determine how the properties of 
galactic and extragalactic jets and QPOs depend on the rotation of the BH.

One of the most tantalizing issues in relativistic
astrophysics is to discover whether
the spin of a BH can be determined from observations.  By far the most
discussed and accepted approach to this problem is through 
the study of the skewed shapes of 
fluorescent iron lines (see Reynolds \& Nowak 2003 for a review). 
Our work provides an independent approach
to this key question for hot, low angular momentum, prograde accretion
flows, as well as for retrograde flows with substantial angular
momentum.

\acknowledgments

PJW is grateful for continuing hospitality at the Department of Astrophysical
Sciences at Princeton University.
PB and PJW are partially supported by RPE funds to PEGA at GSU.
TKD is supported by grants from the KBN.

{}

\end{document}